\documentclass[runningheads]{llncs}
\usepackage[T1]{fontenc}
\usepackage{graphicx}
\usepackage{amssymb}
\usepackage[hidelinks]{hyperref}

\begin{document}
%
\title{Development and Evaluation of an AI-Driven Telemedicine System for Prenatal Healthcare}
%
\titlerunning{An AI System for Prenatal Healthcare}

%


\author{Juan Barrientos\inst{1}\orcidID{0000-0002-4143-8848} \and
Michaelle Pérez\inst{1}\orcidID{0009-0001-7399-4531} \and 
Douglas González\inst{1}\orcidID{0000-0002-8808-1241} 
\and
Favio Reyna\inst{2}\orcidID{0000-0003-1184-5159} \and
Julio Fajardo\inst{1}\orcidID{0000-0003-4034-189X}
\and
Andrea Lara\inst{1}\orcidID{0000-0002-9787-637X}}

\authorrunning{J. Barrientos et al.}
%

\institute{
  BiomedLab, Biomedical Engineering Institute, Universidad Galileo, Guatemala City, Guatemala.\\
  \email{\{juan.barrientos, perezm7,duglasa, julio.fajardo, andrealh\}@galileo.edu}
  \and
  Faculty of Medicine, Universidad Francisco Marroquín, Guatemala City, Guatemala.\\
  \email{Freyna@ufm.edu}
}

\maketitle              
\begin{abstract}
Access to obstetric ultrasound is often limited in low-resource settings, particularly in rural areas of low- and middle-income countries. This work proposes a human-in-the-loop artificial intelligence (AI) system designed to assist midwives in acquiring diagnostically relevant fetal images using blind sweep protocols. The system incorporates a classification model along with a web-based platform for asynchronous specialist reviews. By identifying key frames in blind sweep studies, the AI system allows specialists to concentrate on interpretation rather than having to review entire videos. To evaluate its performance, blind sweep videos captured by a small group of soft-trained midwives using a low-cost Point-of-Care Ultrasound (POCUS) device were analyzed. The system demonstrated promising results in identifying standard fetal planes from sweeps made by non-experts. A field evaluation indicated good usability and a low cognitive workload, suggesting that it has the potential to expand access to prenatal imaging in underserved regions.

\end{abstract}
\section{Introduction}

Access to specialized healthcare services remains a major global challenge. In low- and middle-income countries (LMICs), limited resources and socioeconomic disparities, along with ongoing underinvestment in education and training, reduce the availability of medical care in rural and underserved areas \cite{kruk2018mortality}. While high-income countries average more than three physicians per 1,000 inhabitants, many LMICs remain at or below one physician per 1,000, highlighting a critical gap in healthcare capacity \cite{worldbank_physicians_2024}. Telemedicine has emerged as a vital strategy for bridging healthcare capacity gaps by enabling remote access to specialized care. Since the COVID-19 pandemic, digital health has been increasingly emphasized to expand medical expertise and relieve pressure on local health systems \cite{SaigiRubio2023}. This approach is especially beneficial in regions with significant specialist shortages, where traditional healthcare models are inadequate. In LMICs, tailored telehealth systems have led to improvements in healthcare access \cite{khurram2024sehatkahani,caceres2023telemedicine}. Despite its potential, telemedicine implementation in LMICs remains limited, hindered by factors such as low digital literacy, insufficient training for providers, and sociocultural resistance~\cite{al-samarraieTelemedicineMiddleEastern2020,alnasserTelemedicinePediatricCare2024}. 

Some telemedicine solutions now incorporate artificial intelligence (AI) tools to enhance diagnostic accuracy and scalability in resource-constrained settings. While these AI-based solutions show promising clinical results \cite{liu2022octai}, their adoption in LMICs is still restricted due to a lack of user experience evaluation and cultural relevance. Additionally, most AI tools are developed within narrow clinical contexts, raising concerns about their generalizability across diverse health systems~\cite{assinghvidtLowAdoptionVideo2023,delatorreAnalysisVirtualHealthcare2024}.

Obstetric ultrasound telemedicine tools are vital for prenatal care in rural and underserved areas; however, their use is limited by poor infrastructure, a lack of trained personnel, and insufficient technical support. Most ultrasound machines and AI solutions are designed for hospitals, making them unsuitable for community-based maternal care \cite{paul2023aiultrasound}. Many AI-driven tools provide diagnostic outputs without human oversight, complicating interpretation for healthcare providers, which is critical for effective patient communication \cite{stringer2024diagnostics}. Recent studies suggest human-in-the-loop frameworks as a solution, involving local providers in data collection and trained specialists for interpretation. This approach improves clarity, reduces the cognitive load on non-experts, and fosters trust, ultimately enhancing outcomes in maternal healthcare \cite{vegaOvercomingBarriersUse2025}. In Guatemala, healthcare is centralized in urban areas, restricting access to prenatal care, including obstetric ultrasounds, for rural and Indigenous populations \cite{paho2022hia,who2025guatemala}. Midwives and nurses typically provide primary care in these regions but often lack training and tools for obstetric imaging \cite{martinezMHealthInterventionImprove2018,ramosMobilMonitoringDoppler2024}.

\begin{figure}[b]
\centering
\includegraphics[width=0.75\textwidth]{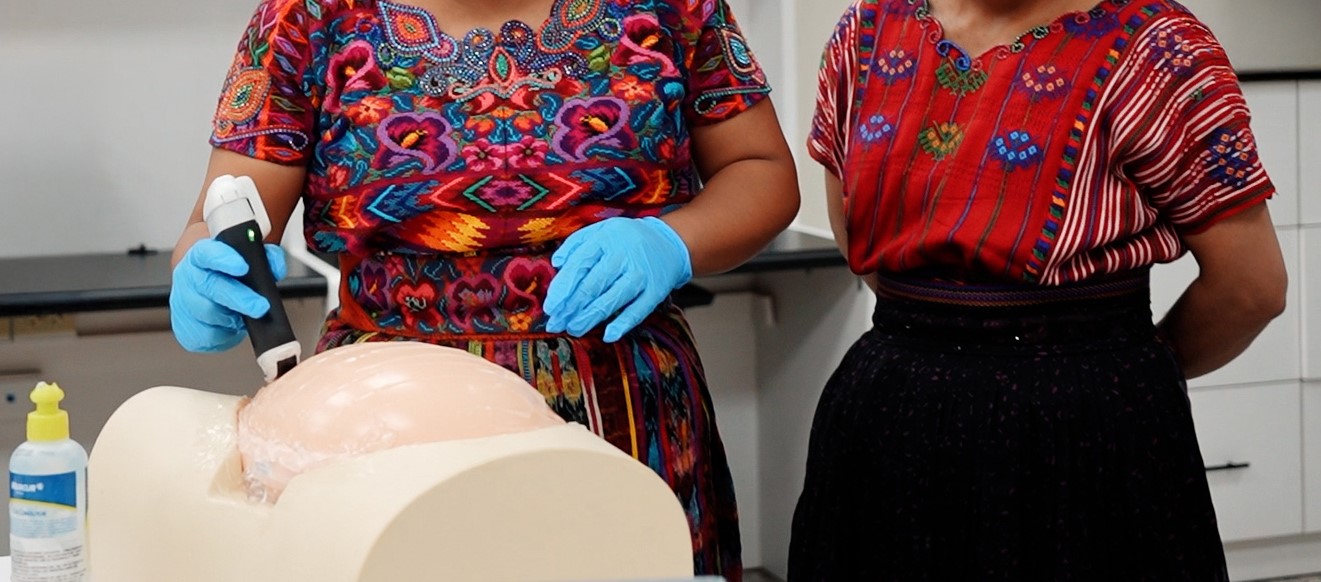} 
\caption{Midwives from rural areas in Guatemala performing blind sweep ultrasound scans using a handheld POCUS device.}
\label{fig:midwives}
\end{figure}

This work presents a pilot study on the NatalIA system conducted with midwives in rural Guatemala, as shown in Fig.~\ref{fig:midwives}. The system integrates portable ultrasound technology, deep learning (DL), a subfield of AI, and community-based care, allowing midwives to perform image acquisition using blind sweep protocols without needing trained personnel \cite{toscanoTestingTelediagnosticObstetric2021}. Collected videos are analyzed by an DL model, which preselects relevant views, allowing remote specialists to focus on interpretation rather than data curation, thus supporting patient follow-up. The study evaluated the acceptance, usability and potential for adoption of the system in community-based maternal care settings. It tested the effectiveness of blind sweep protocols for capturing relevant fetal images and included clinical validation with specialists to assess the model accuracy in identifying standard fetal planes. User perception and cognitive load were analyzed from the midwife's perspective, covering tasks like performing the blind sweep, uploading studies, and interacting with NatalIA interface. Results showed good acceptance among midwives and specialists, indicating that the system is suitable for low-resource environments and has significant potential to connect specialists with rural populations.

\section{Methods}
\subsection{System Architecture}
\begin{figure}[b]
  \centering
  \begin{minipage}[t]{0.48\textwidth}
    \centering
    \includegraphics[width=\textwidth]{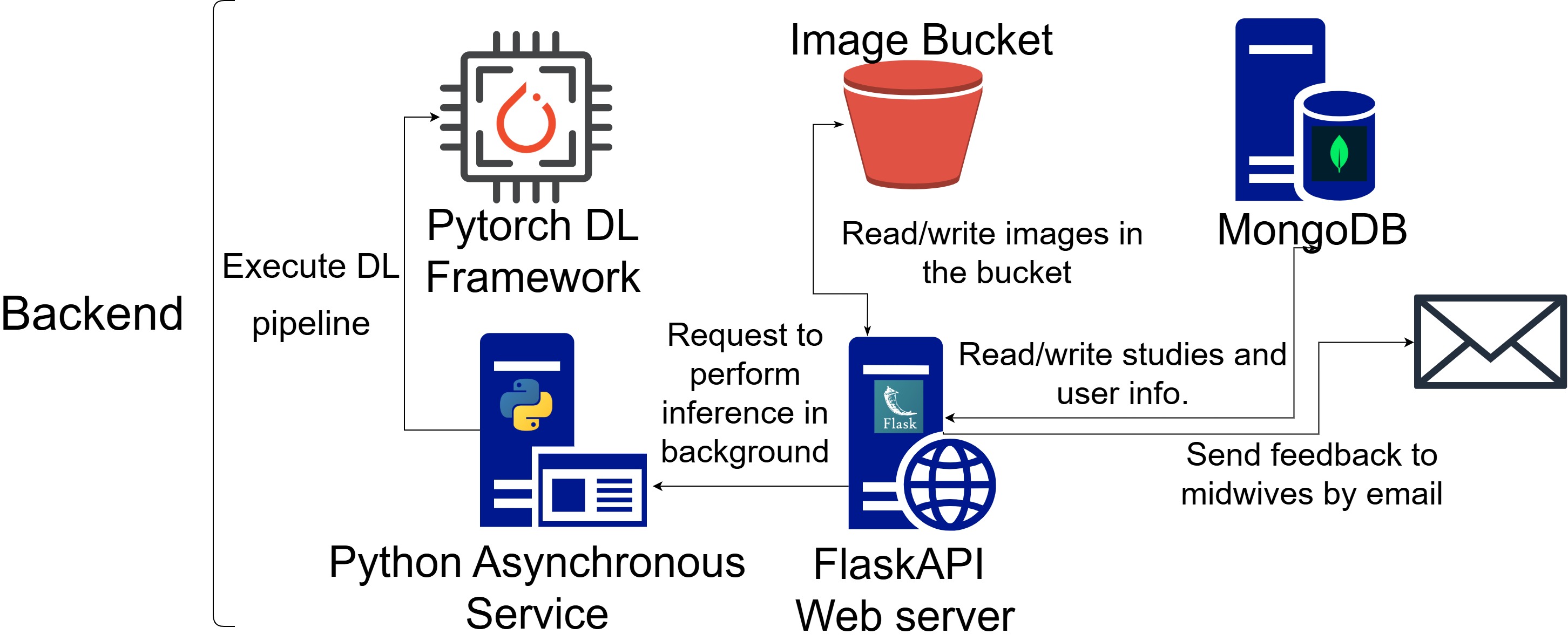}
    \textbf{(a)}
  \end{minipage}
  \hfill
  \begin{minipage}[t]{0.48\textwidth}
    \centering
    \includegraphics[width=\textwidth]{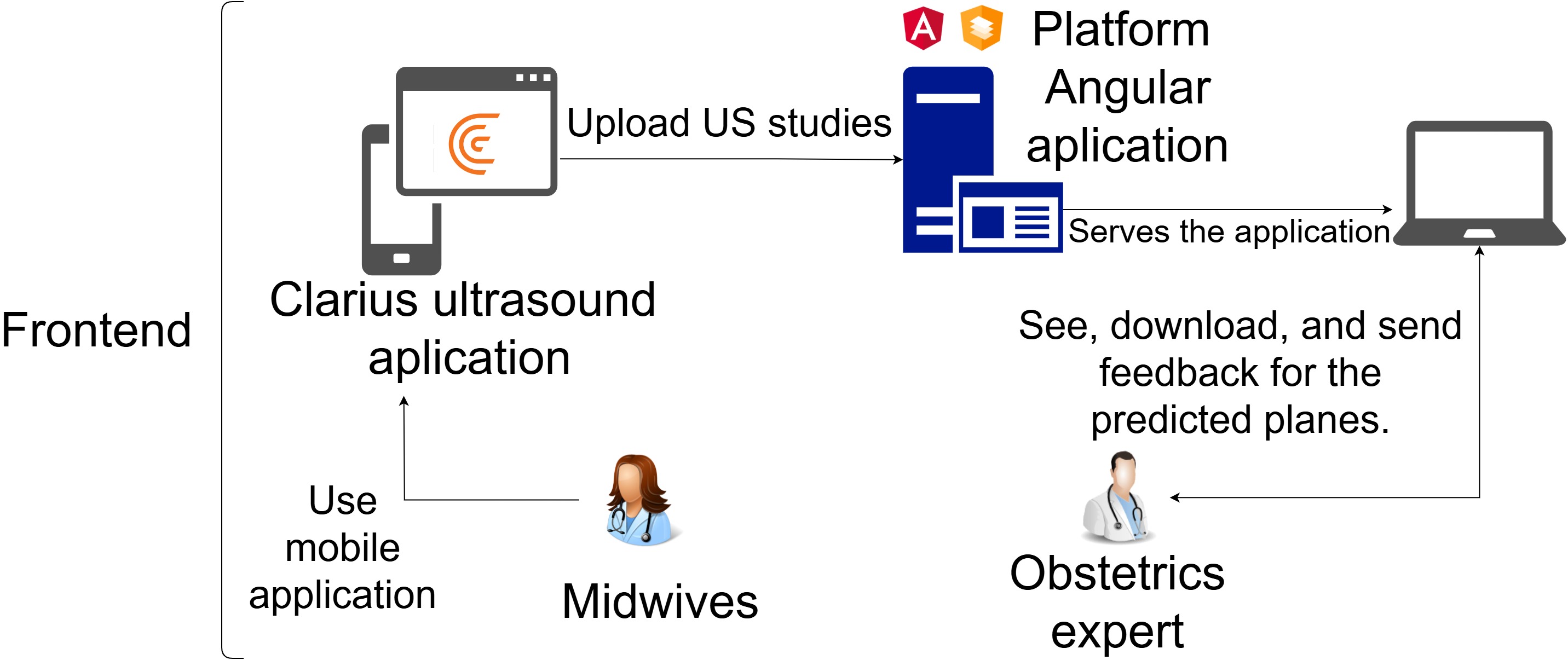}
    \textbf{(b)}
  \end{minipage}
  \caption{System architecture composed of (a) backend and (b) frontend components.}
  \label{fig:sys_arch}
\end{figure}

The proposed system was designed following a client-server architecture with a clear separation between the frontend and backend components, as illustrated in Fig.~\ref{fig:sys_arch}. This web-based design minimizes computational demands on user devices and allows access from any browser without requiring a dedicated application. The frontend was implemented using the Angular framework, providing an interface for midwives to upload ultrasound studies and for specialists to review results. The backend was developed using Flask in Python to manage data processing and communication. Additionally, PyTorch was used to execute the DL model asynchronously, while MongoDB and a local image bucket were used for structured data and image storage, respectively.

Ultrasound videos acquired by midwives using a portable Point-of-Care Ultrasound (POCUS) device and a mobile interface are saved locally and uploaded to the system web platform directly from the same device. Once uploaded, the system processes the studies in the background and automatically identifies standard fetal planes. Ultrasound specialists review these predictions asynchronously via the web interface and may also download the videos if necessary. Finally, clinical feedback is provided to the midwives through the email registered on the platform and also stored within the system for later access, ensuring a traceable and accessible diagnostic workflow.

\subsection{Dataset}
\label{sec:dataset}
\begin{figure}[b]
  \centering
  \includegraphics[width=0.6\textwidth]{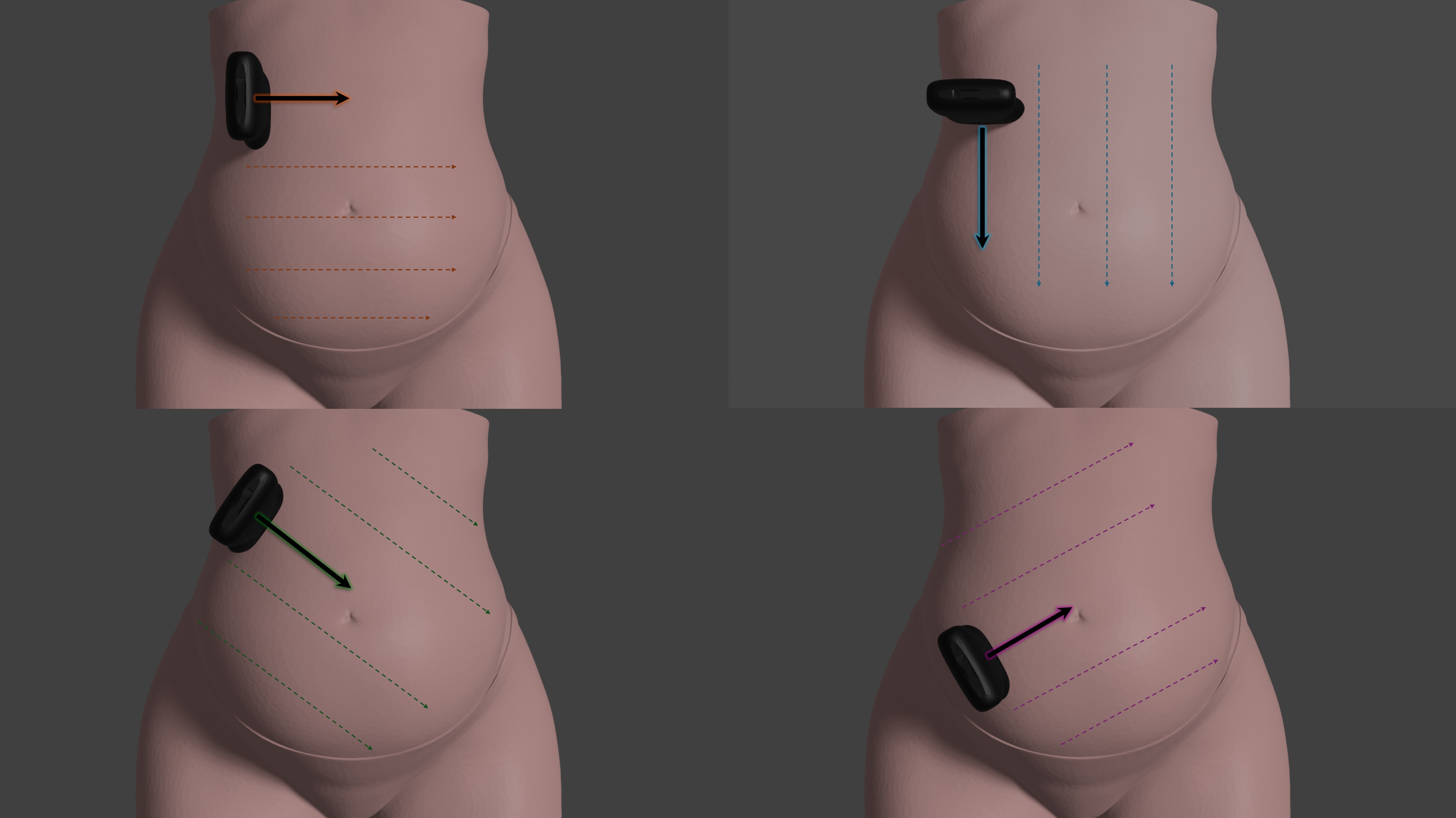}
  \caption{Blind sweep acquisition protocol with  vertical, horizontal, and two diagonal trajectories.}
  \label{fig:blind_sweeps}
\end{figure}

The dataset for this study, NatalIA: PBF-US1 was collected using a US-7a SPACE FAN obstetric phantom (Kyoto Kagaku, Japan) that simulates a 23-week pregnancy~\cite{gonzalez2024natalia}. For its construction, blind sweep ultrasound videos were obtained using a Clarius C3 HD3 handheld POCUS device (Clarius, Canada) under standardized conditions. In addition, forty-five nursing students participated after receiving a one-hour training session on a blind sweep protocol adapted from Toscano et al. \cite{toscanoTestingTelediagnosticObstetric2021}, which includes vertical and horizontal trajectories. We extended the protocol by adding two diagonal sweeps to capture greater anatomical variation. The four types of blind sweeps used in this study are illustrated in Fig~\ref{fig:blind_sweeps}, leading to 90 ultrasound video recordings.

Annotations were performed based on the International Society of Ultrasound in Obstetrics and Gynecology (ISUOG) guidelines for fetal biometry \cite{ISUOG2013FetalHeart,Salomon2011MidTrimester}. The labeling scheme included five standard fetal planes: biparietal, abdominal, heart, spine, and femur, as well as a “no plane” label for frames that did not contain any of the relevant anatomical structures. The dataset comprises 42 biparietal, 63 abdominal, 61 heart, 134 spine, 46 femur plane frames, and 19,061 frames labeled as "no plane".

\subsection{DL Models}
Three different DL architectures were selected for this study in order to extract standard fetal planes from blind sweep ultrasound videos: SonoNet, ResNet-18, and ResNet-50 \cite{7974824,he2015deepresiduallearningimage}. These architectures were chosen due to their relatively lightweight design, which makes them suitable for potential deployment on portable devices commonly used in low-resource or remote environments. Additional frames were incorporated by adding the consecutive frames to the labeled ones. Frames showing high similarity, with both the Structural Similarity Index (SSIM) and normalized cross-correlation exceeding 0.90, were included and assigned the same label as the original frame. This approach expanded the dataset and allowed the model to learn from a wider set of clinically meaningful patterns. After this step, the dataset included 562 abdominal, 264 biparietal, 365 femur, 544 heart, and 1067 spine frames. For the “no plane” class, 30\% of available frames were randomly selected, totaling 4,765 examples. The model was trained and validated using an 80/20 split of the dataset. Its performance was further assessed on blind sweep acquisitions performed by midwives, which specialists independently reviewed to confirm the presence of standard fetal planes.

\section{Experiments and Results}
To evaluate both the technical performance and field usability of the proposed system, a series of experiments were conducted comprising three main components: (1) the training and evaluation of DL models for fetal plane detection, (2) A clinical validation phase where specialists verified the accuracy of the fetal plane detections made by the NatalIA DL model based on blind sweep acquisitions performed by midwives, and (3) a usability study involving midwives and obstetric ultrasound specialists. The following section describes the study design and summarizes the key findings from all three components.

\subsection{DL Model Evaluation}
To evaluate the ability of architectures to classify standard fetal planes in blind sweep acquisitions, we first tested SonoNet, a widely used model for fetal plane detection. This initial experiment used the original pretrained weights without retraining to assess the model’s out-of-domain performance. As expected, results were suboptimal due to domain shift, with an accuracy of 30\% and an F1-score of 39\%, confirming the need for adaptation to the characteristics of blind sweep acquisitions performed by non-experts.

Subsequently, SonoNet was fine-tuned on our dataset using the same training procedure as for other architectures. In parallel, ResNet-18 and ResNet-50 were tested using a transfer learning approach initialized with ImageNet weights. All models were trained with the AdamW optimizer, a learning rate of 0.000001, and a batch size of 64 for 300 epochs. The cross-entropy loss function was used, and data augmentation techniques included random horizontal and vertical flips and rotations up to 45 degrees.

\begin{figure}[ht]
  \centering
  \begin{minipage}[b]{0.32\textwidth}
    \centering
    \includegraphics[width=\textwidth]{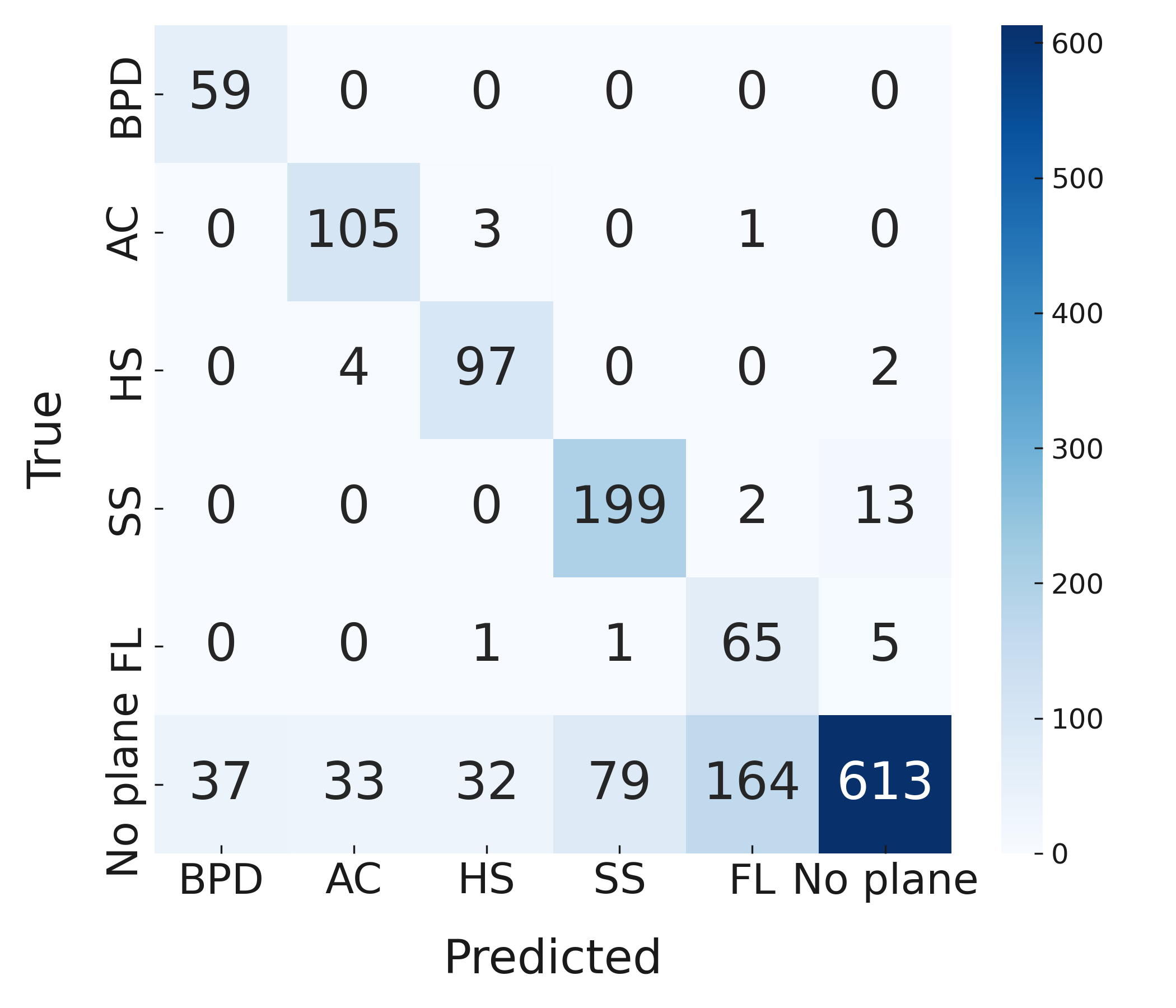}
    \textbf{(a)}
  \end{minipage}
  \hfill
  \begin{minipage}[b]{0.32\textwidth}
    \centering
    \includegraphics[width=\textwidth]{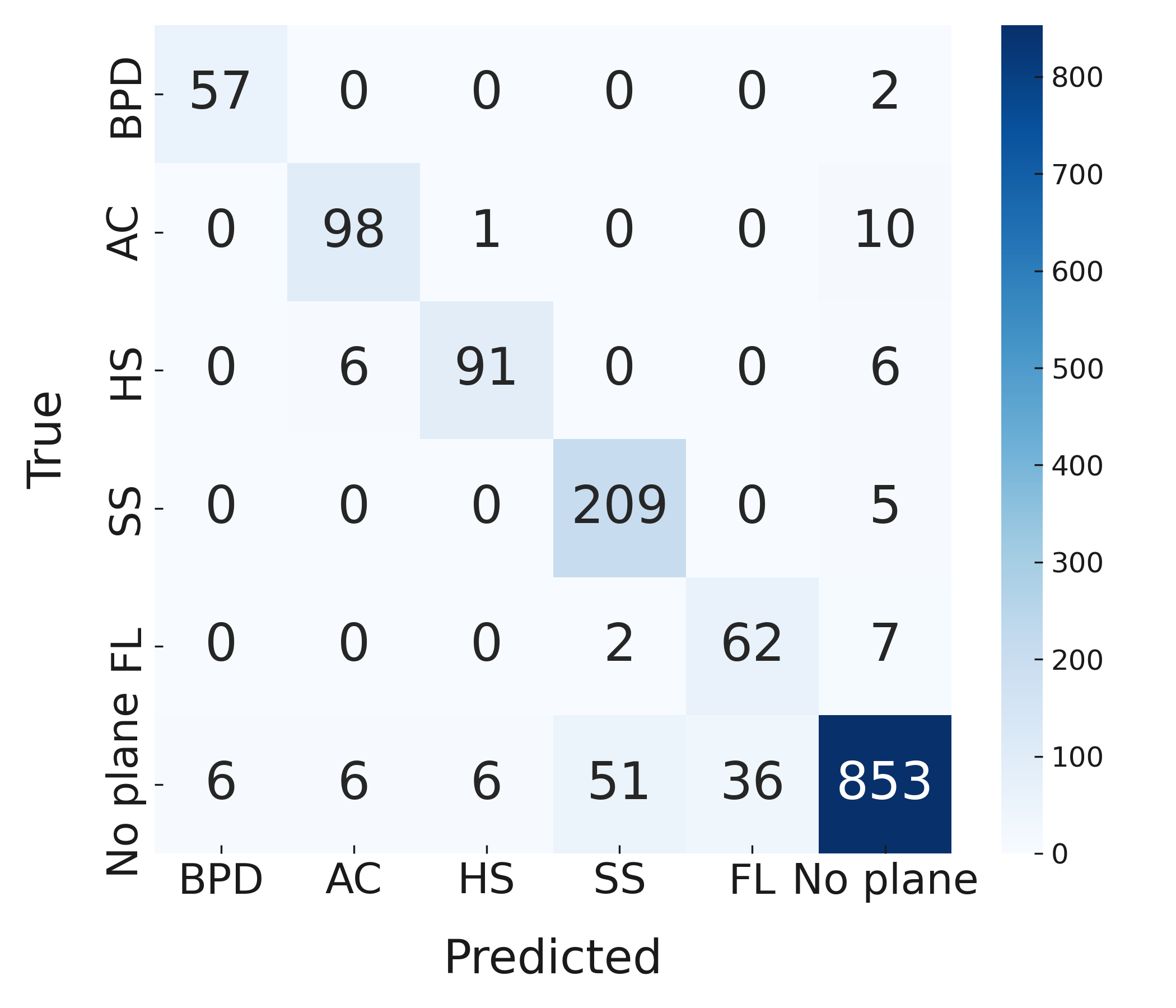}
    \textbf{(b)}
  \end{minipage}
  \hfill
  \begin{minipage}[b]{0.32\textwidth}
    \centering
    \includegraphics[width=\textwidth]{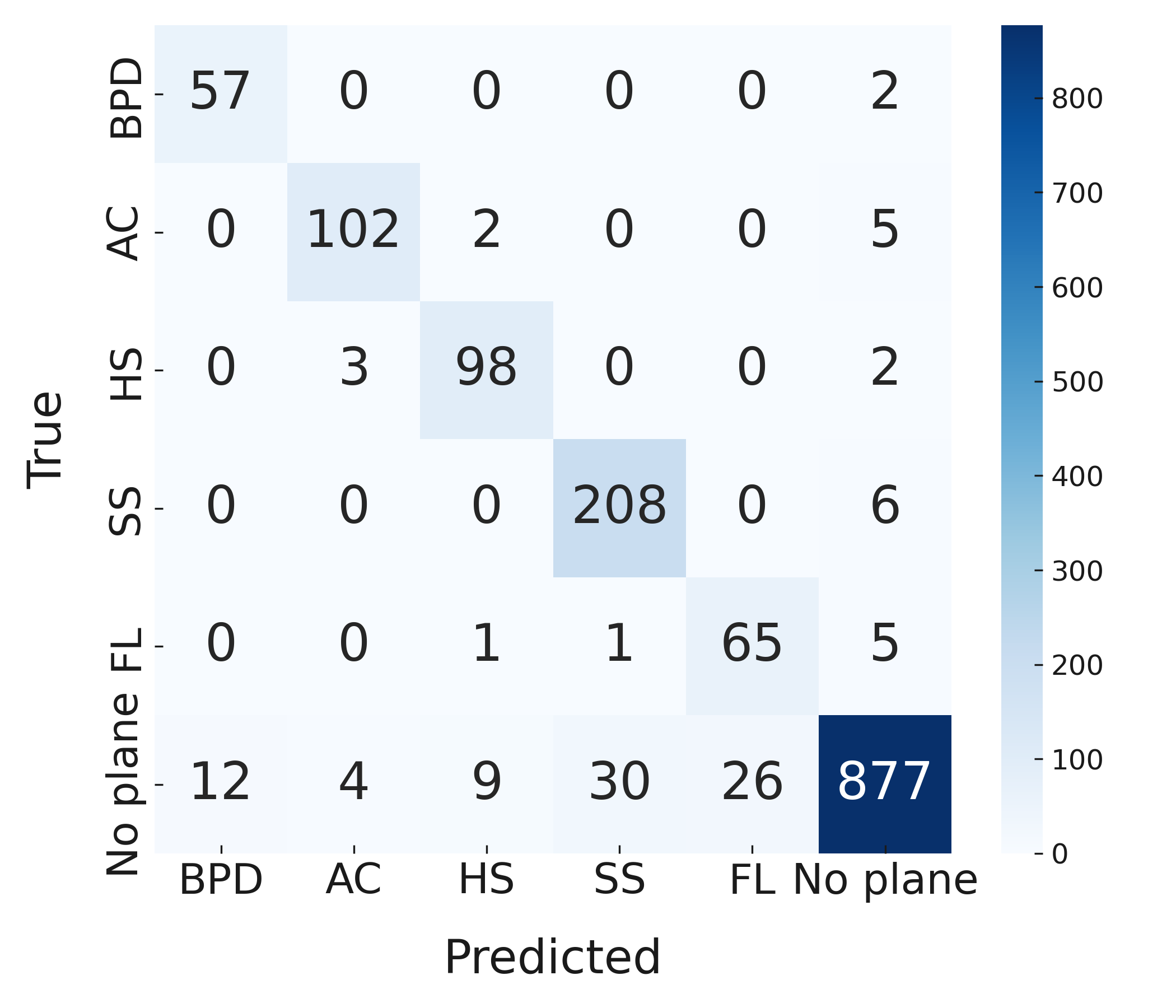}
    \textbf{(c)}
  \end{minipage}
  \caption{Confusion matrices from the transfer learning experiment for the three architectures: (a) SonoNet, (b) ResNet-18, and (c) ResNet-50. Labels indicate fetal planes: BPD (biparietal diameter), AC (abdominal circumference), HS (heart standard view), SS (spine standard view), FL (femur length), and No plane (non-diagnostic frame).
}
  \label{fig:transfer_cm}
\end{figure}

The classification results are presented in the confusion matrices of Fig.~\ref{fig:transfer_cm}. After fine-tuning, \textbf{SonoNet} achieved an accuracy of 75.5\% F1-score of 76.7\%. \textbf{ResNet-18} reached an accuracy of 90.6\% and a F1-score of 90.7\%, while \textbf{ResNet-50} showed the best overall performance with 92.87\% and a F1-score of 93.0\%, demonstrating consistent improvements across fetal plane categories.
\subsection{Clinical Validation of System’s Predictions by Specialists}
A validation test was conducted using a fetal ultrasound phantom to assess the performance in real settings of the system. Eight midwives from rural regions of Guatemala each received a one-hour training session on performing blind sweep acquisitions. After training, each midwife independently performed two blind sweeps on the phantom. They then received a brief introduction to the digital platform and uploaded their recordings. The DL model processed these videos and returned the fetal planes it was able to detect. These outputs were reviewed by ultrasound specialists, who independently evaluated which fetal standard planes were present in the recordings. Table~\ref{tab:midwives_results} shows a side-by-side comparison of the fetal standard planes identified by the DL model and those confirmed by the specialists for each midwife.
\begin{table}[ht]
\centering
\caption{Comparison of fetal planes identified by the system and confirmed by specialists for each midwife.}

\label{tab:midwives_results}
\begin{tabular}{|c|c|c|c|c|c||c|c|c|c|c|}
\hline
\textbf{Midwife} & \multicolumn{5}{c||}{\textbf{NatalIA System}} & \multicolumn{5}{c|}{\textbf{Specialists}} \\
\cline{2-11}
 & AC & BPD & HS & SS & FL & AC & BPD & HS & SS & FL \\
\hline
1 & 6 & 10 & 4 & 1  & 10  & 3 & 4 & 4 & 9 & 5 \\
\hline
2 &  3 & 11 & 0  &  1 & 3 & 2 & 6 & 6 & 4 & 4 \\
\hline
3 & 3 &  9 & 3 & 2 &  8 & 3 & 6 & 6 & 5 & 6 \\
\hline
4 & 4 & 12 & 5 &  2 &  7 & 2 & 6 & 4 & 1 & 3 \\
\hline
5 & 3 & 11 & 3 &  1 &  2 & 3 & 7 & 5 & 1 & 5 \\
\hline
6 & 3 &  7 & 0 &  2 & 12 & 2 & 3 & 3 & 1 & 5 \\
\hline
7 &  4 & 8 & 2 & 4 &  9 & 3 & 4 & 7 & 2 & 5 \\
\hline
8 & 10 &  9 & 1 & 3  & 5  & 2 & 4 & 3 & 0 & 4 \\
\hline
\end{tabular}
\end{table}

\subsection{Field Evaluation Study}
To evaluate the usability and feasibility of the system in real settings, a field study was conducted involving eight midwives from rural regions of Guatemala and three ultrasound specialists. Each participating midwife had between 5 and 15 years of experience in prenatal care, routinely performing procedures such as Leopold maneuvers and monitoring fetal heart rates with Doppler technology. As none of the midwives had previous training or experience with ultrasound techniques, they received prior training on the blind sweep protocol and the web-based platform. The evaluation focused on three main areas: blind sweep usability, system usability, and cognitive workload.
Data for the first two areas were collected through structured questionnaires administered after completing both the training and the actual use of the system in the field. For cognitive workload assessment, the NASA Task Load Index (NASA-TLX) was used to quantify the mental, physical, and temporal demands perceived by the participants during use of the system.

\subsubsection{Blind Sweep Protocol Usability}
After the training session, all midwives were able to perform blind sweeps independently. Most participants reported that the protocol was easy to understand and required only one or two guided attempts to feel confident in its use. All completed two acquisition attempts using the ultrasound phantom. Several participants also mentioned limitations in their local infrastructure, such as unreliable access to electricity or the internet.
\subsubsection{System Usability}
Midwives found the web platform intuitive and easy to navigate. They successfully uploaded studies and accessed feedback using their mobile devices. Ultrasound specialists also reviewed the system interface and confirmed that it supported their clinical workflow. Suggestions for improvement included adding support for Mayan languages and incorporating a real-time help function.
\subsubsection{Cognitive Workload Evaluation}
\begin{figure}[h]
  \centering
  \includegraphics[width=0.7\textwidth]{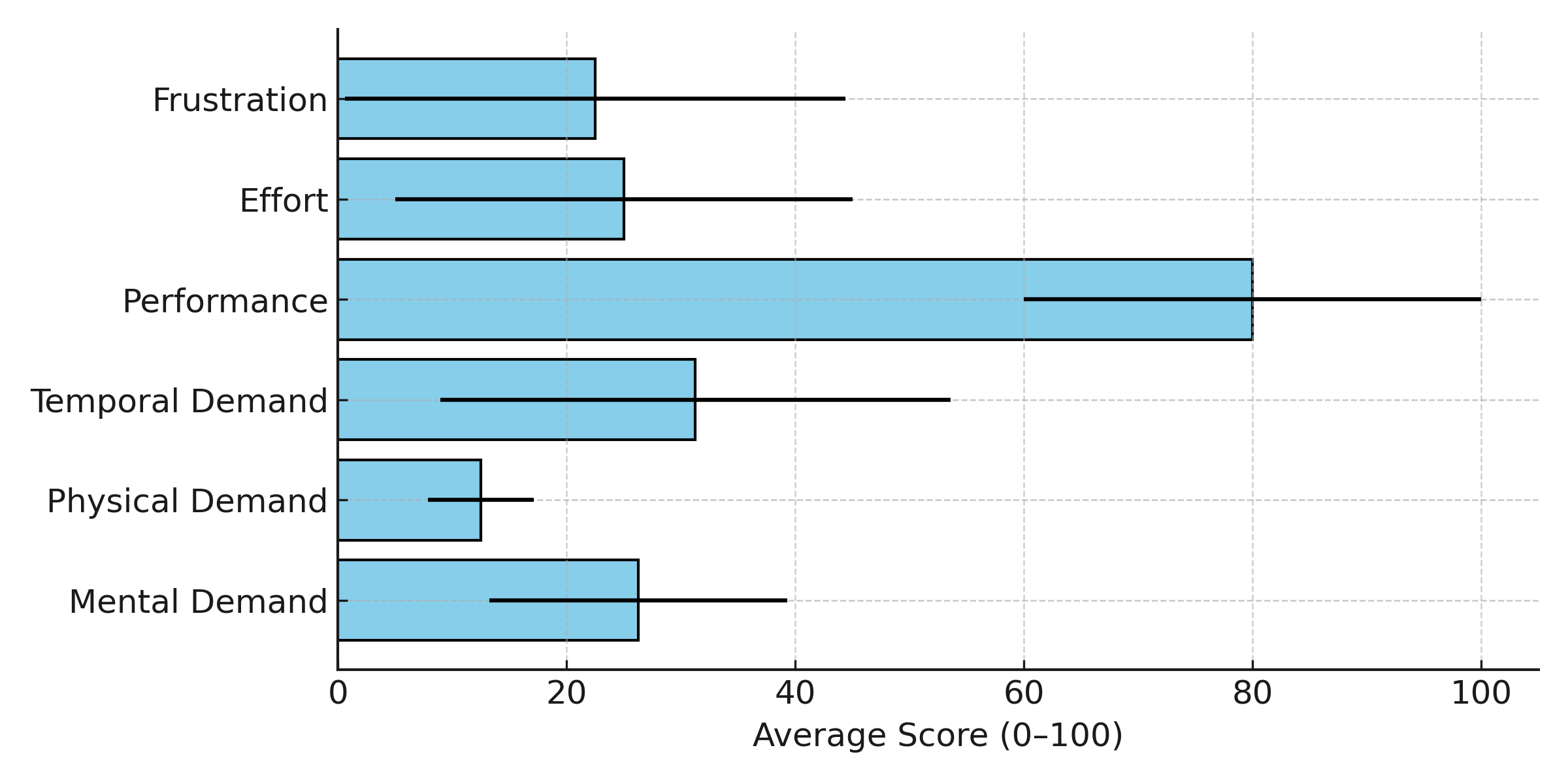}
  \caption{Average scores reported by participants across the six NASA-TLX workload.}
  \label{fig:nasatlx}
\end{figure}

NASA-TLX responses indicated that most participants experienced low to moderate levels of mental demand (M = 26.25, SD = 12.99) and very low physical demand (M = 12.5, SD = 4.14). Time pressure  (M = 31.25, SD = 23.05), frustration (M = 22.5, SD = 20.62), and effort (M = 25.0, SD = 19.15)  were generally rated as low, and the perceived performance score was consistently high (M = 80.0, SD = 21.60). Only a few participants reported isolated increases in workload dimensions.
Fig.~\ref{fig:nasatlx} summarizes the average scores in the six dimensions of the NASA-TLX workload, including standard deviation bars to reflect the variability of the participants. 

\section{Discussion and Conclusion}
This study suggest that a human-in-the-loop DL system can be effectively integrated into community-based maternal care workflows. The proposed system, demonstrated high usability and low cognitive demand when used by midwives with no prior experience in ultrasound. After brief instruction, participants successfully acquired ultrasound studies and interacted with the platform to upload and review findings. The overall process was positively received by both midwives and remote specialists, suggesting strong potential for adoption in low-resource settings, despite that the participants identified ultrasound as the least accessible diagnostic tool due to high equipment costs, limited availability in their communities, and its technical complexity. 

The DL model evaluation showed that the original SonoNet architecture had limited performance under the new data conditions, while the adapted ResNet-based models, especially ResNet-50, achieved significantly better results. These findings highlight the importance of adapting models and using transfer learning when applying DL systems in clinical environments that differ from those used during initial training. The system was able to identify relevant fetal planes from blind sweep videos acquired by users without prior experience, and most of its predictions matched the evaluations made by independent specialists.

Despite the encouraging findings, this study presents several limitations. Data collection was conducted using a phantom model rather than real patients, and the usability assessment involved a small group of midwives.  Moreover, while efforts were made to contextualize the usability findings, a review of the literature revealed a lack of published studies evaluating the usability of tele-ultrasound systems in comparable settings. This absence underscores the novelty and relevance of our work. Future research should include clinical validation with pregnant women, long-term follow-up, and participation from a broader range of primary healthcare providers. Beyond this study, the workflow combines blind sweep acquisition by non-experts, DL-based automatic analysis, and remote specialist review. This approach is valuable where training, time, or equipment are limited. Further evaluation in diverse clinical and cultural settings is needed.

\begin{credits}
    \subsubsection{\ackname} 
    This project was developed with the financial support of the Instituto de Efectividad Clínica y Sanitaria (IECS) trough a grant awarded by the International Development Research Centre (IDRC), Ottawa, Canada. The authors gratefully acknowledge FACISA (Faculty of Health Sciences, Universidad Galileo) for their valuable contributions to the development of the dataset used in this study.”
    
    \subsubsection{\discintname}
    The views and opinions expressed in this document are those of the authors and do not necessarily represent those of Instituto de Efectividad Clínica y Sanitaria (IECS), International Development Research Centre (IDRC), or their governing bodies.
\end{credits}

%
%
\bibliographystyle{splncs04}
\bibliography{references}

\end{document}